\documentclass[a4paper,final]{siamltex}
\usepackage{a4,graphicx,amssymb,latexsym}

\newcommand{\egem}[1]{\emph{#1}}
\newlength{\myfoo}
\title{The Anderson model of localization: a challenge for modern
  eigenvalue methods\thanks{This work was supported by the Deutsche
    Forschungsgemeinschaft within the Sonderforschungsbereich~393
    ``Numerische Simulation auf massiv parallelen Rechnern''.}}

\author{Ulrich Elsner\footnotemark[3] \and
        Volker Mehrmann\footnotemark[3] \and
        Frank Milde\footnotemark[4] \and
        Rudolf A.\ R{\"o}mer\footnotemark[4] \and
        Michael Schreiber\footnotemark[4]}

\begin{document}

\pagestyle{myheadings}
\thispagestyle{plain}
\markboth{U.~ELSNER, V.~MEHRMANN, F.~MILDE, R.~R{\"O}MER AND M.~SCHREIBER}{A
  CHALLENGE FOR MODERN EIGENVALUE METHODS} 

\maketitle
\renewcommand{\thefootnote}{\fnsymbol{footnote}}
\footnotetext[3]{ Fakult{\"a}t f{\"u}r Mathematik, Technische Universit{\"a}t,
                  D-09107 Chemnitz, Germany }
\footnotetext[4]{Institut f{\"u}r Physik, Technische Universit{\"a}t, D-09107
                  Chemnitz, Germany} 
\renewcommand{\thefootnote}{\arabic{footnote}}

\begin{abstract}
  We present a comparative study of the application of modern
  eigenvalue algorithms to an eigenvalue problem arising in quantum
  physics, namely, the computation of a few interior eigenvalues and
  their associated eigenvectors for the large, sparse, real,
  symmetric, and indefinite matrices of the Anderson model of
  localization.  We compare the Lanczos algorithm in the 1987
  implementation of Cullum and Willoughby with the implicitly
  restarted Arnoldi method coupled with polynomial and several
  shift-and-invert convergence accelerators as well as with a sparse
  hybrid tridiagonalization method.  We demonstrate that for our
  problem the Lanczos implementation is faster and more memory
  efficient than the other approaches. This seemingly innocuous
  problem presents a major challenge for all modern eigenvalue
  algorithms.
\end{abstract}




%
%

\section{Introduction}
\label{sec-intro}

In this paper we present a comparative study of the application of
modern eigenvalue algorithms to an eigenvalue problem arising in
quantum physics. The task is to compute a few (5--10) interior
eigenvalues and the associated eigenvectors of a family of structured
large, sparse, real, symmetric, indefinite matrices. The off-diagonal
elements are equal to the off-diagonal elements of the 7--point
central difference approximation to the three-dimensional Poisson
equation on the unit cube with periodic boundary conditions. The
matrices differ from each other only in the diagonal entries, which
are suitably chosen random numbers.

Previously this problem was often solved by using the 1987 Cullum and
Wil\-lough\-by implementation of the Lanczos algorithm
\cite{Culw85a,Culw85b}, in the following called CWI.  But in the last
10 years several new eigenvalue methods have been developed and
implemented as software packages, that seem, at least at first glance,
more appropriate than CWI, see, \egem{e.g.}, the recent survey and
comparison given in \cite{Lehs95}. We apply these new codes to the
described family of matrices and check whether they are faster and
more memory efficient than CWI. To our surprise, none of the tested
codes is consistently better than CWI. As we show below, we find only
a single new code which is at least as fast as CWI. But this code
needs two orders of magnitude more memory than CWI. We therefore
believe that the described family of matrices will present an
important new benchmark example and will hopefully lead to
modifications and improvements for the current methods.

The paper is organized as follows. In \S\,\ref{sec-anderson} we
describe the underlying quantum physics problem, \egem{i.e.}, the
Anderson model of localization, and introduce the parameters used in
our study.  In \S\,\ref{sec-cullum} we briefly review the
Cullum/Willoughby version of the Lanczos method that has been
previously used in the simulations for this model. We then give in
\S\,\ref{sec-modern} a brief survey of more recent eigenvalue methods.
In \S\,\ref{sec-results} we present comparative results for the
different methods and show that CWI is faster and needs less memory
than all other approaches.

%
%

\section{The Anderson model of localization}
\label{sec-anderson}

The Anderson model of localization \cite{And58} is a convenient model
for the investigation of electronic properties of disordered systems.
Although it represents a severe simplification of amorphous materials
and alloys, it has nevertheless become a paradigmatic model and is
currently widely used in the theoretical description of quantum
mechanical effects of disorder such as, \egem{e.g.}, spatial
localization of electronic wave functions with increasing strength of
disorder and the corresponding metal-insulator transitions
\cite{Hof93,Multi,Zara97}.  The quantum mechanical problem is
represented by a Hamilton operator in the form of a real symmetric
matrix $A$ and the quantum mechanical wave functions are simply the
eigenvectors of $A$, \egem{i.e.}, finite vectors $x$ with real
entries.  \egem{E.g.}, for a simple cubic lattice with $M = N \times N
\times N$ sites, we have to solve the eigenvalue equation $Ax =
\lambda x$, which is given in site representation as
\begin{eqnarray}
  x_{i-1,j,k} + x_{i+1,j,k} + x_{i,j-1,k} + x_{i,j+1,k} +
  x_{i,j,k-1} + x_{i,j,k+1} + \varepsilon_{i,j,k} x_{i,j,k} & & 
\label{eq-se-disc}\\
= \mbox{ } \lambda x_{i,j,k}, & & \nonumber
\end{eqnarray}
with $i,j,k$ denoting the cartesian coordinates of a site. The
off--diagonal entries of $A$ correspond to hopping probabilities of
the electrons from one site to a neighboring site.  For simplicity, we
have set them all to unity in (\ref{eq-se-disc}). The disorder is
encoded in the random potential site energies $\varepsilon_{i,j,k}$ on
the diagonal of the matrix $A$.  We consider only the case of
$\varepsilon_{i,j,k}$ being uniformly distributed in the interval
$[-w/2,+w/2]$. This is a common simplification, usually used in the
studies of the Anderson model of localization with typical values of
$w$ ranging from $1$ to $30$. The boundary conditions are usually
taken to be periodic, but hard wall and helical \cite{Helical}
boundary conditions are sometimes also used.  According to the
Gersgorin circle theorem \cite{Golv96} every such matrix $A$ has
eigenvalues in the interval $[-w/2 - 6, +w/2 + 6]$. Possible
generalizations of the Anderson model include anisotropic \cite{Multi}
or even random hopping \cite{Eilmes} and various choices of the
distribution function of the site energies \cite{Gs95}.  However, the
graph of the matrix remains the same.

\begin{figure}
 \begin{center}
  \includegraphics[width=0.9\linewidth]{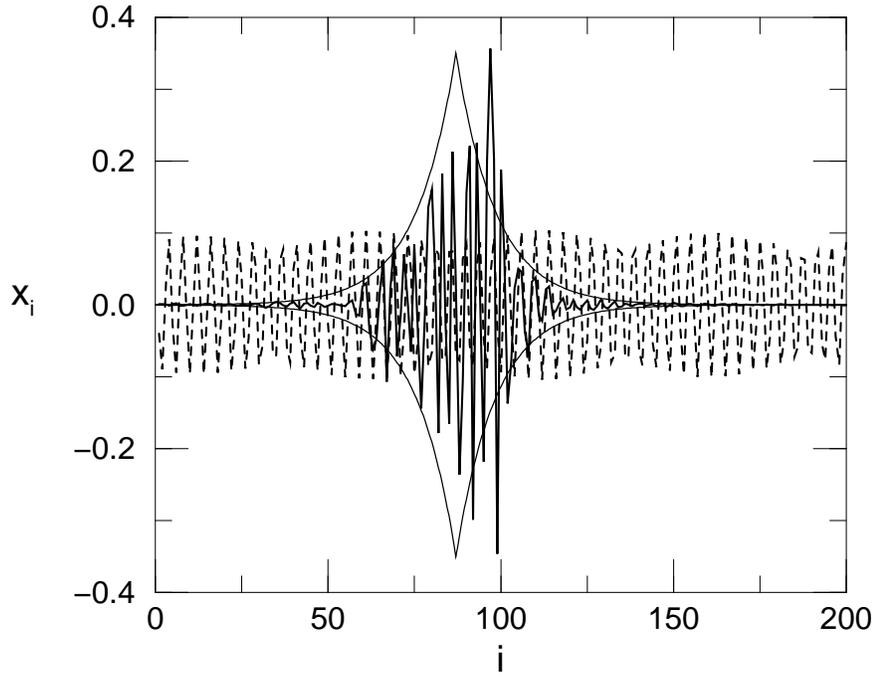}
  \caption{\label{fig-states}
    Extended (dashed line) and localized (thick solid line) eigenstate
    for a single realization of the Anderson model in one spatial
    dimension with $N=200$ sites and periodic boundary conditions. For
    the localized eigenstate, we also show the exponential envelope
    with localization length $l\approx 12$ (thin lines) according to
    \S\,\protect\ref{sec-anderson}.}
 \end{center}
\end{figure}

Although the above matrix seems to be fairly simple, the intrinsic
physics is surprisingly rich. For small disorder ($w \ll 16.5$), the
eigenvectors are extended, \egem{i.e.}, $x_{i,j,k}$ is fluctuating
from site to site but the envelope $|x|$ is approximately a non-zero
constant.  For large disorder ($w \gg 16.5$), all eigenvectors are
localized, \egem{i.e.}, the envelope $|x_n|$ of the $n$th eigenstate
may be approximately written as $\exp[ -|\vec{r} - \vec{r}_n|/l_n(w)
]$ with $\vec{r} = (i,j,k)^T$ and $l_n(w)$ denoting the localization
length of the eigenstate at the specified strength $w$ of the
disorder. In Fig.\ \ref{fig-states}, we show examples of such states
for the Anderson model in one spatial dimension.
Since extended states can contribute to electron transport, whereas
localized states cannot, the Anderson model thus describes a
metal-to-insulator transition: In three-dimensional samples at $w =
w_c \approx 16.5$, the extended states at $\lambda\approx 0$ vanish
and no current can flow. The eigenvector properties are also connected
with the statistical properties of the spectrum $\sigma(A)$ of $A$. In
the extended regime one finds level repulsion, while in the localized
regime the eigenvalues are uncorrelated resulting in level clustering.
These results agree quantitatively with random matrix theory
\cite{Hof93}.  Directly at $w_c$ there is a so-called critical regime
where the eigenvectors are multifractal entities \cite{Multi,Gs91}
showing characteristic fluctuations of the amplitude on all length
scales.  In order to numerically distinguish these three regimes,
namely localized, critical and extended behavior, one needs to (i) go
to extremely large system sizes and (ii) average over many different
realizations of the disorder, \egem{i.e.}, compute eigenvalues or
-vectors for many matrices with different diagonals.

\begin{figure}
 \begin{center}
%
   \includegraphics[width=0.9\textwidth]{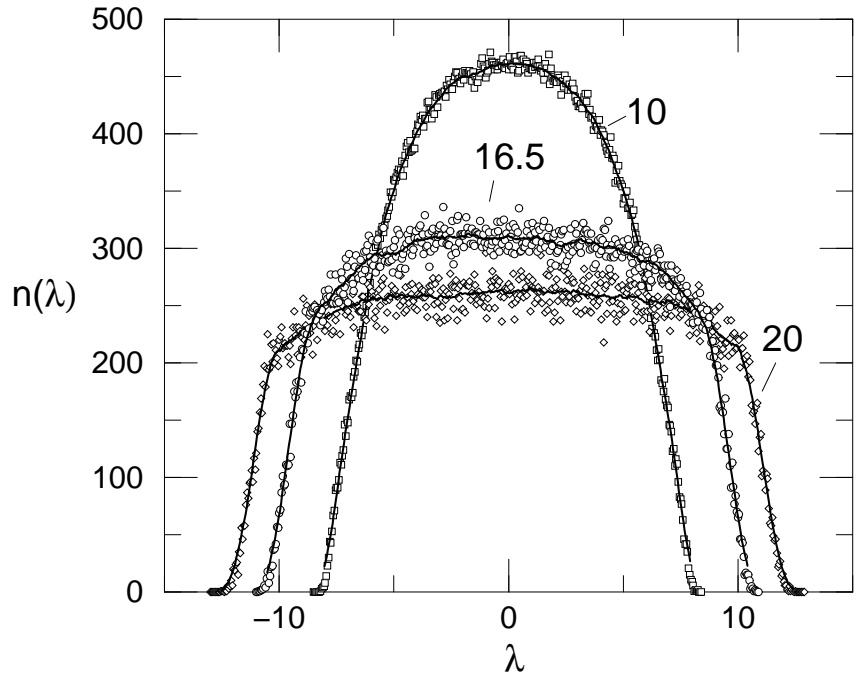}
  \caption{\label{fig-spec}
    Histogram $n(\lambda)$ of eigenvalues for a single system with
    $N^3=48^3$ sites and $w= 10$ ($\protect\Box$), $16.5$
    ($\protect\circ$), and $20$ ($\protect\diamond$). The bin width is
    $0.05$. The lines are obtained by consecutively averaging $20$
    bins. }
 \end{center}
\end{figure}

In the present paper we concentrate on the computation of a few
eigenvalues and corresponding eigenvectors for the physically most
interesting case of critical disorder $w_c$ and in the center of
$\sigma(A)$, \egem{i.e.}, at $\lambda=0$, for system sizes as large as
possible. In Fig.~\ref{fig-spec}, we show a histogram of $\sigma(A)$
for different disorders. Note the high density of states at
$\lambda=0$ in all cases. Therefore we have the further numerical
challenge of distinguishing clearly the eigenstates in this high
density region.

%
%

\section{The Lanczos algorithm and the Cullum/Willoughby implementation}
\label{sec-cullum}

As outlined in the last section, each of the matrices $A$ is sparse,
symmetric and indefinite. Furthermore, the matrix-vector
multiplication $A x$ can be written explicitly as in
(\ref{eq-se-disc}) and is thus easily implemented. An ideal candidate
for an algorithm taking advantage of nearly all these properties is
the Lanczos algorithm \cite{Golv96}. This algorithm iteratively
generates a sequence of orthogonal vectors $v_i$, $i=1,\ldots, K$, such
that $V_K^T A V_K = T_K$, with $V=\{v_1,v_2,\cdots,v_K\}$ and $T_K$ a
symmetric tridiagonal $K\times K$ matrix. One obtains the recursion
\begin{equation}
  \beta_{i+1} v_{i+1} = A v_i - \alpha_i v_i -\beta_i v_{i-1},
\label{eq-lanczos}
\end{equation}
where $\alpha_i=v_i^T A v_i$ and $\beta_{i+1}=v_{i+1} A v_{i}$ are the
diagonal and subdiagonal entries of $T_K$, $v_0=0$ and $v_1$ is an
arbitrary starting vector. For $K=M$ in exact arithmetic this is an
orthogonal transformation to tridiagonal form that needs  $M$
matrix-vector multiplications. The eigenvalues of the tridiagonal
matrix $T_K$, also known as Ritz values, are then simply the
eigenvalues of the matrix $A$ and the associated Ritz vectors yield
the eigenvectors \cite{Culw85a,Culw85b,Golv96,Par80,Saa92}.

In finite precision arithmetic, however, the Lanczos vectors $v_i$
loose their orthogonality after a small number of Lanczos iterations.
Consequently, there appear so called ``spurious'' or ``ghost''
eigenvalues in $\sigma(T_K)$, which do not belong to $\sigma(A)$.

There are several solutions to this problem: total reorthogonalization
of all Lanczos vectors against each other, selective
reorthogonalization \cite{ParS79}, or distinguishing between good and
spurious eigenvalues. While the reorthogonalization leads to an
increase in memory requirements and computing time, since all or
several of the $v_i$ need to be stored and reorthogonalized, the
solution implemented in CWI \cite{Culw85b} uses a simple and highly
successful procedure to identify the spurious eigenvalues, thereby
avoiding reorthogonalization. It thus only uses two Lanczos vectors in
each iteration step and consequently the memory requirements are very
small.  An eigenvalue of $T_K$ is identified as being spurious if it
is also an eigenvalue of the matrix $T'_K$ which is constructed by
deleting the first row and column of $T_K$.  Still, the good
eigenvalues produced may not yet have converged properly for a given
$K$. So we further use the fact that good eigenvalues will be
replicated in $\sigma(T_K)$ if $K$ is large enough. We only accept
eigenvalues as being good eigenvalues after they have been replicated
at least once in $\sigma(T_K)$.  Hence we usually need at least $K
\geq 2 M$. Finally, in order to obtain the eigenvectors corresponding
to these good eigenvalues of $A$, all Lanczos vectors must be computed
a second time, again doubling the computational effort.

The convergence of the Lanczos algorithm is very fast for the
eigenvalues close to $\min\,\sigma(A)$ and $\max\,\sigma(A)$. This is
especially true if these eigenvalues are well separated.  However, for
eigenvalues in the interior of $\sigma(A)$ and for eigenvalues which
are not well separated the convergence is slow. Furthermore, the
tridiagonal matrix $T_K$ becomes very large for an iteration in the
interior of $\sigma(A)$.  Nevertheless, the CWI has been used to study
the Anderson model of localization even at $\lambda=0$ successfully
for years \cite{Schreib90,Gs91,Hof93,Gs95,Multi,Zara97} and
eigenstates for matrices with $N=100$ can be obtained within a few
weeks of computing time \cite{Zara_priv_97}.

We also remark that most of the computational effort in the Lanczos
algorithm is spent on the iteration of (\ref{eq-lanczos}),
\egem{i.e.}, on matrix-vector multiplications and vector additions.
These can be easily parallelized and thus the CWI is well suited for
parallel architectures. For example, the eigenspectra presented in
Fig.\ \ref{fig-spec} have been obtained by such a parallel version of
CWI running for about 60 hours for each realization using 16
processors of a Parsytec GCC Power Plus.

%
%

\section{Modern approaches}
\label{sec-modern}

Lately there has been much progress in eigenvalue methods mostly
concentrating on non-symmetric matrices. We refer to \cite{Lehs95} for
a recent survey. The symmetric problem is usually assumed to be taken
care of implicitly.  But although our symmetric eigenvalue problem is
well-conditioned \cite{Golv96}, the fact that the eigenvalues are
clustered in the neighborhood of $\lambda=0$, our region of interest,
creates difficulties for all numerical methods.  Promising choices for
possible replacements of the Cullum and Willoughby approach are the
implicitly restarted Arnoldi method \cite{Lehs95} and the hybrid
tridiagonalization (HTD) algorithm of Cavers \cite{Cav94}. Another new
approach is the Jacobi-Davidson method \cite{BooVMR94}. In the
following we will pay special attention to the Arnoldi approach, since
it allows the easy use of the shift-and-invert technique. We expect
this to overcome the above mentioned clustering problem at $\lambda=0$.

\subsection{Modifying the eigenproblem}

The problem of slow convergence in the interior of $\sigma(A)$ can be
overcome by computing eigenvalues and eigenvectors for a modified
eigenvalue problem $f(A)x = f(\lambda)x$. The function $f$ is chosen
such that the desired point $\lambda$ in $\sigma$ is mapped onto or
close to the minimum or maximum of $\sigma(f)$.  Furthermore, one
should choose $f$, such that $\sigma(f)$ has well separated
eigenvalues at $\min\,\sigma(f)$ and $\max\,\sigma(f)$.

Among the many possible choices for $f(A)$, we shall consider in the
following: \linebreak[3]
(i) \emph{polynomial convergence accelerators}, where $f(A)$ is
chosen as a polynomial which has its maximum (or minimum) at
$\lambda$. This moves $\lambda$  to $\max\,\sigma(f(A))$ (or
$\min\,\sigma(f(A))$ resp.). We remark that occasionally these
convergence accelerators are somewhat misleadingly called
preconditioners. \linebreak[3]
(ii) \emph{shift-and-invert} with $f(A)=(A-\lambda I)^{-1}$ and $I$
the $M\times M$ identity matrix. This choice of $f$ requires the
additional solution of a linear system with $A-\lambda I$ in each step
of the eigenvalue iteration \cite{Golv96,Saa92}. For the solution of
this linear system there are again two alternatives:\linebreak[3]
(ii.a)~\emph{direct sparse solvers} for $A-\lambda I$. Unfortunately,
not many direct solvers exist which can make efficient use of 
the sparseness for indefinite problems.\linebreak[3]
(ii.b)~\emph{iterative solvers} using only matrix-vector multiplications
with $A-\lambda I$. The iterative methods promise to make large matrix
sizes possible, since they benefit in an optimal way from sparsity.
Memory requirements and computational cost of a matrix-vector
multiplication are proportional to the number of non-zeros and
therefore proportional to $M$ for our present problem.  However, we
note that in our case $(A-\lambda I)$ is indefinite which will lead to
slow convergence for most iterative solvers.  Since the convergence of
the iterative solver is dominated by the condition number of the
linear system, one may employ preconditioners to accelerate its
convergence \cite{Golv96}.

All these approaches result in a competition between smaller numbers
of Lanczos or Arnoldi iterations and increased costs for each such
iteration step. For this reason it is not a priori clear whether they
will indeed give a net reduction in computation time.

\subsection{The implicitly restarted Arnoldi method}

In a recent comparison \cite{Lehs95} of different Arnoldi based
packages {\sc ArPack} \cite{Lehsy} was found to be the fastest and
most reliable of the compared codes.

When applied to symmetric eigenvalue problems, the main difference
between the techniques in \textsc{ArPack} based on the Arnoldi
iteration and the Lanczos iteration is an implicit restart technique.
The Arnoldi method stores a number of Ritz vectors produced by the
iterations and after a small number of steps initiates a restart which
uses an implicit QR--algorithm for the small eigenvalue problem to
create a new starting vector and to maintain orthogonality among the
Ritz vectors. In contrast to the Lanczos algorithm more vectors have
to be stored but spurious eigenvalues are avoided.

The {\sc ArPack} implementation further allows the easy use of
additional acceleration methods such as polynomial convergence
acceleration and shift-and-invert as outlined above. So {\sc ArPack}
is probably the best choice for a replacement of CWI.

\subsection{Other approaches}
\label{sec:other}

We have also studied the use of the HTD method \cite{Cav94} and the
Jacobi/Davidson method \cite{BooVMR94}. The HTD method is a direct
tridiagonalization method specifically designed for sparse matrices.
This makes it an interesting approach for our purposes. However, it is
not explicitly designed to compute only few interior eigenvalues and
associated eigenvectors.

The Jacobi/Davidson method appears to be another promising future
direction if it can be properly accelerated. The current version is
designed for complex unsymmetric problems and to report comparative
results would not be fair to this interesting new development. We
intend to further study this method in the future.

%
%

\section{Results}
\label{sec-results}

After a short discussion of the specific implementations and
parameters, we now present the results of our comparison. The tests
are performed on Hewlett-Packard HP9000 735/125 workstations for
$N^3\le 24^3$ and on a HP9000 K460 with the fast PA8000 processor for
$N^3\ge 24^3$. The latter machine allows us to use up to 1.9 GB RAM
and is about 3.5 times faster. In order to obtain a fair comparison we
always require that the eigenresidual of the computed
eigenvalue/eigenvector pair satisfies $|A x - \lambda x|\le 10^{-8}$.
The CPU times have been measured using the {\sc Unix} {\tt time}
command of the {\tt tcsh} shell. Even for the largest system sizes
considered, we have usually taken at least $5$ different realizations
of disorder and averaged the resulting CPU times. Sometimes, when the
CPU times for a given algorithm fluctuate widely, we report the range
of times instead of a simple average. We remark that the use of {\tt
  time} introduces a further uncertainty into the results such that we
always have an error of about $10\%$.  The random number generators
used are {\tt ran2} from \cite{Numrec} and the {\tt rand} command from
{\sc Matlab}.

\subsection{The standard approach}

For our particular problem we can reach $M=80^3=512000$ with CPU
times of about two weeks on the K460 machine using CWI. However,
keeping in mind the configurational averaging necessitated by the
underlying physical problem, a reasonable upper limit for the matrix
size is $M=50^3=125000$.

In Table~\ref{tab-times-intedge} we show the CPU times obtained for
CWI in the center and at the edge of $\sigma(A)$. Note that the
computing times are nearly independent of the disorder parameter $W$,
but, as expected, CWI is much faster at the edges of $\sigma(A)$ than
at $\lambda=0$.  In Table~\ref{tab-times-poly}, we show the results
for CWI at $\lambda=0$ in dependence on the matrix size $M$.


\begin{table}
\footnotesize
\begin{center}
\settowidth{\myfoo}{101.5}
\begin{tabular}[h]{|r|r@{.}l|r@{}l|c|r@{}l|} \hline
 \multicolumn{1}{|c|}{$M$\rule[-1ex]{0mm}{3.5ex}}&
 \multicolumn{2}{c|}{$W$}  &\multicolumn{2}{c|}{CWI}  & 
 {\sc ArPack} & \multicolumn{2}{c|}{HTD} \\
\hline \hline
1000  & 10&0 &  2&.4 & \phantom{0}240 -- 2200\phantom{0}    &  22&  \\
1000  & 16&5 &  2&.5 & \phantom{0}230 -- 1400\phantom{0}    &  22&  \\
1000  & 20&0 &  2&.4 & \phantom{0}140 -- 1300\phantom{0}    &  23&  \\
1728  & 10&0 &  7&.9 &           1700 -- 12000              & 170&    \\
1728  & 16&5 &  7&.8 & \phantom{0}410 -- 12000              & 170&    \\
1728  & 20&0 &  7&.6 &           1100 -- 20000              & 160&    \\
4096  & 10&0 & 43&   &                                      &2600&    \\
4096  & 16&5 & 40&   &                                      &2500&    \\
4096  & 20&0 & 40&   &                                      &2500&    \\
\hline \hline
1000  & 10&0 &  0&.71& \makebox[\myfoo][r]{0.78}    &  22&  \\
1000  & 16&5 &  0&.77& \makebox[\myfoo][r]{0.85}    &  22&  \\
1000  & 20&0 &  0&.80& \makebox[\myfoo][r]{0.89}    &  23&  \\
1728  & 10&0 &  0&.94& \makebox[\myfoo][r]{1.5}    & 170&    \\
1728  & 16&5 &  1&.0 & \makebox[\myfoo][r]{1.7}    & 170&    \\
1728  & 20&0 &  1&.1 & \makebox[\myfoo][r]{1.8}    & 160&    \\
13824 & 10&0 &  9&.4 & \makebox[\myfoo][r]{57\phantom{.55}}&  &  \\
13824 & 16&5 &  9&.2 & \makebox[\myfoo][r]{57\phantom{.55}}&  &  \\
13824 & 20&0 &  9&.3 & \makebox[\myfoo][r]{71\phantom{.55}}&  &  \\
\hline
\end{tabular}
\smallskip
\end{center}
 \caption{
   CPU times in seconds to compute 5 eigenvectors with CWI, {\sc
     ArPack}, and HTD. The upper part of the table corresponds to
   eigenvalues in the interior of $\sigma(A)$ at $\lambda\approx 0$,
   the lower part corresponds to the 5 largest eigenvalues.}
\label{tab-times-intedge}
\end{table}

\begin{table}
\footnotesize
\begin{center}
\begin{tabular}[h]{|r|r@{}l|r|r@{}r|r@{}l|r|} \hline
&
&
& CWI+
&&
& \multicolumn{2}{c|}{{\sc ArPack\rule{0mm}{2.5ex}}+}
& \\
  \multicolumn{1}{|c|}{\raisebox{1.5ex}[-1.5ex]{\quad $M$}}
& \multicolumn{2}{c|}{\raisebox{1.5ex}[-1.5ex]{CWI}}
& conv.~acc.
& \multicolumn{2}{c|}{\raisebox{1.5ex}[-1.5ex]{\sc ArPack}}
& \multicolumn{2}{c|}{conv.~acc.}
& \raisebox{1.5ex}[-1.5ex]{HTD} \\ 

\hline \hline
1000   & 2&.5    &5.6              & 230 -- & 1400  & 4&.9           & 22   \\ 
1728   & 7&.8    &11\phantom{.0}   & 410 -- & 12000 & 10&            & 170  \\ 
4096   & 40&     &59\phantom{.0}   &        &       & 66&            & 2500 \\
13824  & 770&    &1700\phantom{.0} &        &       & 1700&          &      \\ 
\hline
13824  & 220&    &                 &        &       & 550       &    &      \\ 
27000  & 1000&   &                 &        &       &\quad 3500 &    &      \\
91125  & 20000&  &                 &        &       &           &    &      \\ 
110592 & 35000&  &                 &        &       &           &    &      \\ \hline
\end{tabular}
\smallskip
\end{center}
\caption{
  CPU times in seconds to compute at $w=16.5$ the eigenvectors
  corresponding to the 5 eigenvalues closest to $\lambda=0$ with CWI,
  CWI with Chebyshev-polynomial acceleration, {\sc ArPack} in normal
  mode, {\sc ArPack} with Chebyshev-polynomial acceleration and HTD
  for various matrix sizes $M$. The CPU times in the upper (lower)
  part of the table have been measured on the HP 735 (HP K460).}
\label{tab-times-poly}
\end{table}

\begin{table}
\footnotesize
\begin{center}
\begin{tabular}[h]{|c|r|r|r|} \hline

& \multicolumn{3}{c|}{NCV \rule[-1.5ex]{0mm}{4ex}} \\

  \raisebox{1.5ex}[-1.5ex]{\hspace{2em}}    
& 90& 110& 130 \\

\hline \hline
1      & 2577& 4337& 9455\\
2      & 2011& 2267& 3455\\
3      & 4190& 1935& 3635\\
4      &  411&  755& 1204\\
5      & 2435& 1811&11620\\
6      & 8188& 4704& 2506\\
\hline
\end{tabular}
\smallskip
\end{center}
 \caption{
   CPU times in seconds to compute 5 eigenvectors with {\sc ArPack}
   for 6 different diagonals and 3 choices of NCV for $M=1728$ and
   $w=16.5$ at $\lambda=0$.}
\label{tab-times-ncv-interior}
\end{table}

\begin{table}
\footnotesize
\begin{center}
\begin{tabular}[h]{|c|r|r|r|r|r|r|} \hline
       
& \multicolumn{6}{c|}{NCV \rule[-1.5ex]{0mm}{4ex}}\\

\hspace{2em}
& \multicolumn{2}{c|}{$w=10$}
& \multicolumn{2}{c|}{$w=16.5$}
& \multicolumn{2}{c|}{$w=20$}\\
& 15 & 20 & 15 & 20 & 15 & 20 \\
\hline\hline
1      & 41& 44 & 63 & 62 & 56 & 58 \\
2      & 65& 65 & 43 & 47 & 49 & 55 \\
3      & 57& 60 & 54 & 64 & 44 & 47 \\
4      & 62& 62 & 44 & 49 & 49 & 55 \\
5      & 72& 76 & 72 & 68 & 76 & 71 \\
6      & 44& 45 & 56 & 62 & 87 & 85 \\
7      & 81& 70 & 72 & 88 &108 & 90 \\
8      & 52& 60 & 49 & 51 & 70 & 66 \\
9      & 42& 48 & 58 & 68 & 99 & 75 \\
\hline
\end{tabular}

\smallskip
\end{center}
 \caption{
   CPU times in seconds to compute the eigenvectors corresponding to
   the 5 largest eigenvalues of $A$ with {\sc ArPack} for 9 different
   diagonals and 2 choices of NCV with $M=13824$.}
\label{tab-times-ncv-edge}
\end{table}

In the {\sc ArPack} implementation of Lehoucq et al.\ \cite{Lehsy} one
has to set the parameter NCV which is the largest number of basis
vectors that will be used in the implicitly restarted Arnoldi process.
In normal mode and in the interior of $\sigma(A)$ we find that the
actual value of NCV heavily influences computing time as shown in
Table \ref{tab-times-ncv-interior}. This dependence on NCV becomes
less pronounced for eigenvalues close to $\min\,\sigma(A)$ and
$\max\,\sigma(A)$ as shown in Table \ref{tab-times-ncv-edge}. However,
since we do not know of any strategy to choose NCV optimally, this is
a severe restriction of {\sc ArPack}.  Furthermore, as shown in
Table~\ref{tab-times-intedge}, {\sc ArPack}, working in normal mode,
is much slower than CWI both in the center and at the edge of
$\sigma(A)$. It becomes too slow for practical use already at
$M=12^3=1728$ as shown in Table~\ref{tab-times-poly}.

In Tables \ref{tab-times-intedge} and \ref{tab-times-poly} we also
include CPU times for the HTD method. Note that we only show the CPU
times needed to transform $A$ to tridiagonal form. Nevertheless, we
find that HTD is much slower than CWI. We remark that when we use CWI
to compute the full spectrum as in Fig.\ \ref{fig-spec}, it is still
faster than HTD except for small system sizes $M \leq 12^3=1728$.

\subsection{Polynomial convergence acceleration}

\begin{figure}
 \begin{center}
%
  \includegraphics[width=0.9\textwidth]{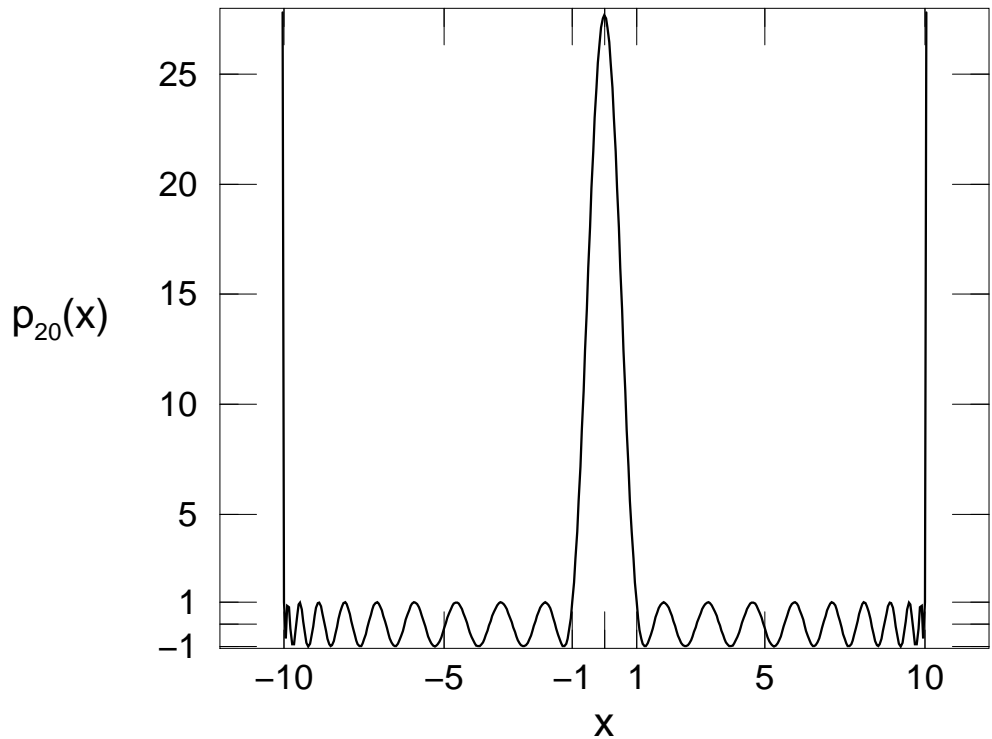}
  \caption{\label{fig-cheb}
    Chebyshev polynomial $p_{20}(x)$ with $x_1=1$ and $x_2=10$.}
 \end{center}
\end{figure}

As outlined above, polynomial convergence acceleration is usually a
convenient choice to speed up the computation of eigenvalues and
-vectors corresponding to a small region of $\sigma(A)$. Here, we test
a polynomial provided by D.\ Sorensen, one of the authors of {\sc
  ArPack}, and C. Sun \cite{Dan}. It is based on a Chebyshev-type
polynomial given by the following recursion:

\begin{equation}
  \label{Cheb}
  \begin{array}{lcl}
    p_1(x)&=&1\\
    p_2(x)&=&a+b x^2\\
    p_{n+1}(x)&=&2(a+b x^2) p_n - p_{n-1}\\
  \end{array}
\end{equation}
where $a=(x_1^2+x_2^2)/(x_1^2-x_2^2)$ and $b=2/(x_2^2-x_1^2)$.  Also,
$p_n$ is symmetric with a local maximum $p_n(0)>1$ at zero,
$|p_n(x)|\le 1$ in the intervals $[x_1,x_2]$ and $[-x_2,-x_1]$, and
$p_n$ grows rapidly for $|x|>x_2$ as shown, \egem{e.g.}, in
Fig.~\ref{fig-cheb} for $n=20$.  In general, one would like to have
$x_1$ and $x_2$ chosen automatically in order to obtain a suitable
function $f$ as described in \S\,\ref{sec-modern}. In all our present
calculations we use $n=50$ with $(x_1)^2= 0.005$ and $(x_2)^2 = 1.1\,
[ \max\, \sigma(A) ]^2$ according to \cite{Dan}.
This polynomial convergence acceleration speeds up {\sc ArPack}
immensely as one can see in Table~\ref{tab-times-poly}. Since
$\lambda=0$ is now mapped to $\max\, \sigma(p_{50}(A))$, the actual
value of NCV is less important. We found NCV~$=50$ to be a good choice
to make the execution times faster, although this requires more
memory.  Still CPU times are about a factor of two larger than for CWI
without any convergence acceleration.  Unfortunately, CWI itself is
not made faster by the use of this accelerator as also shown in Table
\ref{tab-times-poly}.  Although the number of Lanczos vectors needed
to achieve convergence is reduced remarkably, the additional
computational effort now required for every Lanczos step becomes very
large. At the end one needs even slightly more matrix-vector
multiplications than without convergence acceleration.

\subsection{Shift-and-invert with direct solvers}

We now discuss the use of the shift-and-invert mode of {\sc ArPack}
together with a direct solver for the linear system $(A - \lambda I) y
= b$. We first note that although our matrix $A$ is symmetric, it is
not positive definite and thus we cannot use a sparse Cholesky
decomposition.  Unfortunately, there are only few packages available
for sparse symmetric indefinite problems \cite{Superlu}. Therefore we
also investigated several packages for general sparse matrices.

{\sc Meschach} \cite{Meschach} is a freely available mathematical
package written in C. There are three sparse factorization methods
implemented in {\sc Meschach}: Cholesky, LU, and
Bunch-Kaufmann-Parlett (BKP). Cholesky factorization does not work due
to the indefiniteness of $A$. LU and BKP are supposed to take
advantage of the sparseness of our problem. However, we find that they
have huge memory requirements of the order of $M^2$ as shown in Table
\ref{tab-memory}. So they are inapplicable for large system sizes.
And even for small systems they turn out to be much too slow as shown
in Table~\ref{tab-times-si-direct}.

\begin{table}
\footnotesize
\begin{center}
\begin{tabular}[h]{|r|r@{}l|r@{}l|r@{}l|r@{}l|c|r@{}l|r@{}l|r@{}l|}\hline

&&      
& \multicolumn{2}{c|}{{\sc ArPack}+ }
&&
& \multicolumn{9}{c|} {{\sc ArPack\rule[-1ex]{0mm}{3.5ex}}+} \\

 \multicolumn{1}{|c|}{\raisebox{1.5ex}[-1.5ex]{\quad $M$}}
& \multicolumn{2}{c|}{\raisebox{1.5ex}[-1.5ex]{CWI}}
& \multicolumn{2}{c|}{conv.~acc.}
& \multicolumn{2}{c|}{\raisebox{1.5ex}[-1.5ex]{HTD}}
& \multicolumn{2}{c|}{LU}  
& BKP   
& \multicolumn{2}{c|}{SuperLU} 
& \multicolumn{2}{c|}{MA27} 
& \multicolumn{2}{c|}{MA27+HB} \\
\hline\hline
1000   &  0&.24 &        0&.6   &  1&.4  &  7&.4 & 10& 7&.8   &    0&.8&        0&.4\\
1728   &  0&.43 &        1&.2   &  6&.7  & 18&   & 22&19&     &    2&.2&        0&.9\\
4096   &  1&.0  &        2&.3   & 13&    & 80&   & 81&\qquad92&&   8&.6&        2&.9\\
13824  &  3&.4  &        6&.6   &   &    &   &   &   &  &     &   70&  &       22&\\
27000  &  6&.5  &\qquad 15&     &   &    &   &   &   &  &     &  200&  &       68&\\
91125  & 22&    &         &     &   &    &   &   &   &  &     & 1300&  &      500&\\
110592 & 27&    &         &     &   &    &   &   &   &  &     & 2600&  &\qquad600&\\
\hline
\end{tabular}
\smallskip
\end{center}
\caption{
  Memory requirements in MB to compute at $w=16.5$ the eigenvectors
  corresponding to the 5 eigenvalues closest to $\lambda=0$ for the
  different diagonalizers (Names as in the text, HB indicates hard
  wall boundary conditions)}
\label{tab-memory}
\end{table}

The {\sc Harwell} Subroutine Library \cite{Hsl} contains the sparse
symmetric indefinite solver MA27. As shown in Table
\ref{tab-times-si-direct}, {\sc ArPack} with MA27 is about as fast as
CWI. In fact, MA27 seems to become faster than CWI for $M \ge 45^3 =
91125$.  Unfortunately we could not test this because of the huge
memory requirements of MA27 as shown in Table~\ref{tab-memory}.

\begin{table}
\footnotesize
\begin{center}
\begin{tabular}[h]{|r|r@{}l|r|r|r@{}l|r@{}l|r@{}l|}\hline
&&   
& \multicolumn{8}{c|} {{\sc ArPack\rule[-1ex]{0mm}{3.5ex}}+} \\

  \multicolumn{1}{|c|}{\raisebox{1.5ex}[-1.5ex]{\quad $M$}}     
& \multicolumn{2}{c|}{\raisebox{1.5ex}[-1.5ex]{CWI}}
& LU  
& BKP
& \multicolumn{2}{c|}{SuperLU} 
& \multicolumn{2}{c|}{MA27} 
& \multicolumn{2}{c|}{MA27+HB}\\

\hline\hline
1000     &     2&.5 & 39    &74   &  8&.8&       1&.3 &   0      &.88 \\
1728     &     7&.8 & 150   &300  & 28&  &       5&.0 &   2      &.0 \\ 
4096     &    40&   & 1200  &1900 &\quad220&  & 39&   &   9      &.8 \\
13824    &   770&   &       &     &   &  &     740&   & 140      &   \\
\hline              
13824    &   220&   &       &     &   &  &   260&   &  58      &   \\
27000    &  1000&   &       &     &   &  &  1300&   & 250      &   \\
91125    & 20000&   &       &     &   &  & 19000&   &\qquad4900&   \\
\hline
\end{tabular}
\smallskip
\end{center}
\caption{
  CPU times in seconds to compute at $w=16.5$ the eigenvectors
  corresponding to the 5 eigenvalues closest to $\lambda=0$ with
  shift-and-invert {\sc ArPack} and different direct solvers.  For
  easier comparison, we also include CWI. The CPU times in the upper
  (lower) part of the table have been measured on the HP 735 (HP
  K460).}
\label{tab-times-si-direct}
\end{table}

A noteworthy fact is that MA27 is much better for hard wall boundary
conditions (HB). This can be explained by the fact that the bandwidth
of the matrix is $\mathcal{O}(N^2)$ instead of $\mathcal{O}(N^3)$ as
for periodic boundary conditions.  However, on physical grounds, a
calculation with HB is expected to be much more influenced by the
finite size of the cubes considered. So although we can obtain larger
system sizes here, the results for the interesting physical quantities
may not be as reliable. Nevertheless, {\sc ArPack} with MA27 for
matrices with HB is faster than CWI for matrices with periodic
boundary conditions. MA27 with HB is also faster than CWI with HB,
since for CWI there is only a negligible difference in computing time
between HB and periodic boundary conditions coming from the
matrix-vector multiplication.

{\sc SuperLU} is a package by Demmel \emph{et al.}~\cite{Superlu}
doing a sparse LU decomposition.  Compared with CWI and MA27, {\sc
  SuperLU} is much slower as shown in Table~\ref{tab-times-si-direct}.
Furthermore, it needs about one order of magnitude more memory than
MA27 as shown in Table \ref{tab-memory}. {\sc SuperLU} allows the
input of different preorderings in addition to the default
minimum-degree ordering.  We have tested a symmetric minimum degree
ordering from the {\sc Matlab} program and a nested dissection
ordering computed by the Chaco package \cite{Chaco}.  For some choices
of diagonals we derive small savings in run time and/or memory but
these are not consistent, \egem{i.e.}, the same kind of ordering
speeds up the program for one choice of $N$ and slows it down for
$N+1$.

\subsection{Shift-and-invert with iterative solvers}

Considering the recent advances in iterative solvers, we initially
hoped that {\sc ArPack} in shift-and-invert mode coupled with a modern
iterative method for the solution of linear systems would be quite
efficient. As we will show below, this is not the case.

The quasi-minimal-residual (QMR) technique should be one of the best
iterative solvers for symmetric matrices that works using only
matrix-vector multiplications if no preconditioning is used
\cite{Fren91}.  However, as shown in Fig.~\ref{fig-spec}, our matrices
are indefinite with a nearly symmetric eigenvalue distribution around
zero. This results in a very bad iteration count of about $2 M$ for
the solution of a single linear system of size $M$.  The times and
iteration numbers from three variants implemented in {\sc QMRPack}
\cite{Qmrpack}, namely, QMR based on three-term Lanczos with and
without look-ahead (QMRL/QMRX) and QMR based on coupled two-term
Lanczos without look-ahead (CPX) are not very different as shown in
Table \ref{tab-times-si-iterative}. For all three methods the
iteration count is rather high. Consequently, we find that {\sc
  ArPack} in shift-and-invert mode coupled with {\sc QMRPack} as
iterative solver is about 20 times slower than CWI. The {\sc ArPack}
input parameter NCV was set to 15.  We also find that the
implementation of QMR based on coupled two-term Lanczos with
look-ahead does not converge for larger systems within 50000
iterations.

\begin{table}
\footnotesize
\begin{center}
\begin{tabular}[h]{|r|r@{}l|r|r|r|} \hline

& &
& \multicolumn{3}{c|}{{\rule{0mm}{2.5ex}\sc ArPack}+} \\

  \multicolumn{1}{|c|}{\raisebox{1.5ex}[-1.5ex]{\quad $M$}}
& \multicolumn{2}{c|}{\raisebox{1.5ex}[-1.5ex]{CWI}}
& \multicolumn{1}{c|}{QMRX}
& \multicolumn{1}{c|}{QMRL}
& \multicolumn{1}{c|}{CPX}  \\

\hline\hline
1000     & 2&.5  & 68   &  93&  85\\
1728     & 7&.8  & 239  & 320& 330\\
4096     & 40&   & 1600 & &2300\\
13824    & 770&  & 35000& &\\
\hline              
13824    &  220& & 12000& &\\
27000    & 1000& &      & &\\
91125    &20000& &      & &\\
\hline
\end{tabular}
\smallskip
\end{center}
\caption{
  CPU times in seconds to compute at $w=16.5$ the eigenvectors
  corresponding to the eigenvalues closest to $\lambda=0$ with
  shift-and-invert {\sc ArPack} and the iterative solvers from {\sc
    QMRPack}. For easier comparison, we also include CWI. The CPU
  times in the upper (lower) part of the table have been measured on
  the HP 735 (HP K460).}
\label{tab-times-si-iterative}
\end{table}

In order to check if other iterative methods are perhaps more
efficient than QMR for our family of matrices, we have also tried
several such iterative solvers using the {\sc Matlab} programming
environment.  In addition to QMR, we have considered the
conjugate-gradient-squared method (CGS) \cite{son89}, the
BiConjugate-Gradient method (BiCG) \cite{fle74}, its stabilized
variant (Bi-CGSTAB) \cite{dv92} and the generalized-minimal-residual
(GMRES(k)) method \cite{saas86}.
Furthermore, several general purpose preconditioners \cite{Templates},
\egem{i.e.}, the Jacobi (jac) preconditioner, the ILU(0)
preconditioner and also the three main diagonals as the
preconditioning matrix (tri) have been tested.  Since the performance
of {\sc Matlab} programs cannot directly be compared to compiled
programs, we only give the iteration count of each algorithm.  One
such iteration requires at least one matrix-vector multiplication and
two inner products and is thus at least as expensive as one Lanczos
step.  We always use the built-in implementations of these algorithms
as in {\sc Matlab} v5.1.

In Table~\ref{tab-counts-si-iterative}, we show results obtained for
various matrix sizes $M$. The ranges reflect the variations
corresponding to 12 different realizations of disorder on the diagonal
of the matrices. Note that for the same $M$ we use the same 12
diagonals for all algorithms. We always choose $x_0 = 0$ as initial
vector. The iteration count represents the number of iterations needed
to solve the matrix equation $A x = y$ up to a relative accuracy of
$10^{-8}$.  We always stop the algorithms if after $2M$ iterations
this accuracy has not been achieved.  For practical restart values $k
\le 200$, GMRES($k$) does not converge at all within our iteration
limit. With no restarts, GMRES needed $M$ or slightly less iterations.
But note that both memory and computing-time requirements for $M$
steps of pure GMRES exceed those of a non-sparse direct solver. 


None of the tested preconditioners is consistently effective. The
Jacobi preconditioner in fact increases the iteration count most of
the time.  The ILU(0) preconditioner returns a singular matrix and
consequently appears inapplicable. The tridiagonal preconditioner is
more effective, in some cases reducing the iteration count by up to
50\%.  But again there are examples where it fails to do anything. We
remark that in general the iteration count is consistent with the
results from {\sc QMRPack}. To sum up, we find that all of these
iterative algorithms do not perform better that the QMR algorithm and
consequently are no real alternative.

\begin{table}
\footnotesize
\begin{center}
\newcommand{\nc}{\emph{n.c.}}
\newcommand{\phan}{\phantom{1}}
\begin{tabular}[h]{|r|c|c|c|c|} \hline
$M$   &QMR\rule[-1ex]{0mm}{3.5ex}    & CGS   & Bi-CG       & Bi-CGSTAB\\
 \hline\hline
 512  & \phan796 -- 999& \phan843 -- \nc &\phan809 -- 980 &1006 -- \nc \\
1000  &     1705 -- \nc&     1762 -- \nc &    1701 -- \nc &        \nc \\
1728  &     2918 -- \nc&             \nc &    2932 -- \nc &        \nc \\
2744  &     4809 -- \nc&             \nc &    4775 -- \nc &        \nc \\
4096  &     7270 -- \nc&             \nc &    7401 -- \nc &        \nc \\
\hline
\multicolumn{5}{c}{ }\\ \hline
$M$   & GMRES\rule[-1ex]{0mm}{3.5ex}& GMRES($5N$)  & QMR+jac     &  QMR+tri  \\ \hline\hline
 512  & 511-- 512 & \nc       & 813 -- \nc    & \phan295 -- 569 \\
1000  & 995--1000 & \nc       & \nc           &     1389 -- \nc \\
1728  & 1723--1728& \nc       & \nc           &          \nc      \\
2744  & 2736--2744& \nc       & \nc           &          \nc      \\
4096  &           & \nc       & \nc           &          \nc      \\
\hline
\end{tabular}
\smallskip
\end{center}
\caption{
  Number of iterations needed in {\sc Matlab} in order to solve the
  linear system $A y = b$. The abbreviations for the different
  algorithms are explained in the text. The runs are aborted when the
  number of iterations is more than $2M$. This case of no convergence
  is indicated by
  ``n.c.''.}
\label{tab-counts-si-iterative}
\end{table}

Another idea is to work with the matrix $A^2$ instead of $A$. Since it
is symmetric and positive definite, we can now use the conjugate
gradient method.  But this squares the condition number of the linear
system, which is already usually very large for $A$ \cite{Hac94}.
Hence more effort has to be invested into the development of a good
preconditioner. We find for our matrices that while the iteration
count is in general a bit less than for the methods mentioned above
and the preconditioners are more consistently effective we still need
of the order of $M$ steps with at least two matrix-vector
multiplications for the solution of one linear system.  And since the
shift-and-invert {\sc ArPack} still needs to solve several linear
systems, all the iterative methods working on $A^2$ were no match for
CWI.

\section{Summary}
\label{sec-concl}

We have tested several modern methods to compute a few inner
eigenvectors of a very large sparse matrix corresponding to the
Anderson model of localization motivated within theoretical physics.
Particularly the implicitly restarted Arnoldi method in connection
with polynomial convergence acceleration and in shift-and-invert mode
with several direct and iterative solvers for systems of linear
equations is compared to the Cullum/Willoughby implementation of the
Lanczos method. Despite the recent progress in linear system solvers
we find all considered modern methods to be inapplicable for very
large system sizes, because either the computation times or the memory
requirements are much to large. To sum up, we find that CWI Lanczos is
currently still the most efficient method for the matrix type we are
interested in. We emphasize that the CWI Lanczos, with our slight
modifications as outlined in \S\,\ref{sec-cullum}, is a reliable tool
for our problem. In particular, the problem of spurious eigenvalues
which plague the original Lanczos algorithm, can be handled safely.

Since large scale diagonalizations are widely used in theoretical
physics --- and also theoretical chemistry \cite{Chemlanczos} --- we
would be happy to learn about any algorithm that does better than CWI
for our matrices.  We are especially interested in a preconditioner
for the iterative methods which is suitably adapted to our problem.
Certainly improved direct methods for our matrix type are also of
great importance. We hope to have convinced the reader that it may be
worthwhile to rethink seemingly easy problems like the present
eigenproblem for real and symmetric matrices.

\subsection*{Acknowledgments} 
We thank Iain Duff for supplying us with {\sc Harwell}'s MA27 routine
and Dan Sorensen and Chao Sun for their polynomial convergence
accelerator.

\bibliographystyle{siam}
\bibliography{diag}

\end{document}